\documentstyle[amsfonts,epsf,aps]{revtex}
\newcommand{\ee}{\end{equation}}
\newcommand{\be}{\begin{equation}}
\newcommand{\eea}{\end{eqnarray}}
\newcommand{\bea}{\begin{eqnarray}}
\newcommand{\ea}{\end{array}}
\newcommand{\ba}{\begin{array}}
\begin{document}
%\title{
\begin{center}
{\bf
Hamiltonian model of heat conductivity and Fourier law}
\vskip 3mm
%\author{
{\rm Christian Gruber}
\vskip 1mm
%\address{
{\it Institut de Th\'{e}orie
des Ph\'enom\`enes Physiques, Ecole Polytechnique
F\'ed\'erale de Lausanne, 

CH-1015 Lausanne, Switzerland}
%}
\vskip 3mm
%\author
{\rm Annick Lesne}\vskip 1mm
 % \address{
{\it Laboratoire de Physique Th\'eorique 
des Liquides, Case  121, 
Universit\'e Pierre et Marie Curie,  

4 Place Jussieu, 75252
 Paris Cedex 05,
France}
%}
\vskip 3mm
%\maketitle
\end{center}

\vskip 10mm
\begin{center}
\begin{minipage}{120mm}
{\small
We investigate the stationary nonequilibrium states of a quasi
one-dimensional system of heavy particles whose interaction
is mediated by purely elastic collisions with light particles,
in contact at the boundary with two heat baths with fixed
temperatures $T^+$ and $T^-$.
It is shown that Fourier law is satisfied with a thermal
conductivity proportional to $\sqrt{T(x)}$ where $T(x)$
is the local temperature. Entropy flux and entropy production are 
also investigated.

}
\end{minipage}
\end{center}



\section{Introduction}

Although the Fourier's law  $j_Q=-\;\kappa\;{\rm grad}\,T$,
relating the macroscopic heat flux $j_Q$ to the temperature gradient
${\rm grad}\,T$, 
% through a linear response coefficient $\kappa$
%(heat conductivity, possibly varying in space),
 has been introduced  almost two centuries  ago \cite{Fourier}, 
its microscopic basis is still an open issue.
Actually,
% till recently, 
its status has long been  purely phenomenological 
and its only justification  was then its accurate and faithful agreement
with observations in numerous experimental instances.
Following the advances of nonequilibrium statistical mechanics,
several microscopic models has been recently 
 introduced aiming at {\it deducing}
the Fourier's law from microscopic principles and  relating the 
conductivity to the microscopic parameters of the system.
In  particular, the main issue
is to determine
  what are the
minimal requirements to get Fourier's law  \cite{dhar}  \cite{garrido}
\cite{rondoni} \cite{casati} \cite{joel} \cite{JPE}.

The present study belongs to this line of research. It exploits
recent results \cite{GPL}\cite{GPL3} about the so-called ``adiabatic
piston problem'' to device a spatially extended, quasi-one-dimensional 
system, in contact at its ends with two heat baths at different temperatures
($T^-$ on the left and $T^+$ on the right).
It is composed of an array of $K$ ``pistons'' (heavy point particles)
separated by compartments filled with $\Delta N$ non-interacting
light particles of mass $m$.
The actual interactions are limited to elastic
collisions between light particles and adjacent pistons:
successive pistons are thus coupled through their interactions
with the same fluid, while at the same time successive
fluid compartments are coupled through their interactions with
the same intermediary piston.
%Such an array is somehow reminiscent
% of coupled map systems, [refs?].
The differing boundary  temperatures force the system out of equilibrium,
and one of the issues tackled in  this paper is to determine
the temperature profile in the system.
A key point, discussed in \cite{GP} and \cite{GPL3}, is that as soon
as temperatures differ on each side of a piston,
 its stochastic motion
induces an heat transport (whereas the pistons are adiabatic when fixed).
%basically because collision rates
%between particles and piston depend on the fluid temperature.
Our model 
%(an array of pistons separated with fluid compartments)
is simple enough  to determine a consistent stationary state (with no drift) 
%and associated 
for any given heat flux; it thus sheds light on the controversial issue of heat
 conductivity in 1-dimension 
\cite{lepri}.
The derivation is performed analytically, in the frame of a perturbation 
approach developed in \cite{GPL} with small parameter $m/\Delta M$
where $m$ is the mass of fluid particles and $\Delta M$ the mass of one 
piston.
% It sheds light on the controversial issue of heat
% conductivity in dimension 1
%\cite{lepri}.
%  (here, no divergence when the system length
%increases at fixed $N$, $p$ and $j_Q$).
Such an array is somehow reminiscent of chains of masses linked by springs,
much studied since the pioneering work of Fermi, 
Pasta and Ulam \cite{FPU} \cite{JPE99}.
The masses correspond here to the pistons and the springs to the fluid 
compartments. But rather than being the nonlinearities, the key ingredient
at the origin of transport in our setting will appear to be nonequilibrium fluctuations. Our model can also be related to the general class of models
introduced by Eckmann and Young \cite{JPE} with the pistons playing the role of the ``Energy Storing Devices'', and the fluid particles the role of the ``Conducting Agents''.

The paper is organized as follows.
In Section II, we describe the stationary state and heat flux for a single
piston surrounded with two fluids at different temperatures.
In Section III, we investigate the fluid separating two pistons
 and the indirect coupling of the pistons  that it achieves.
In Section IV, we then bridge these two sets of results to get
the behavior and heat conductivity of the above-mentioned array.
In Section V, the Fourier's law is recovered and discussed. In Section VI,
  we give explicit relations for entropy production and dissipation,
before ending the paper with a final
Section VII devoted to a discussion of the scope of our results.

\section{Single piston stationary state with heat flux  (and no drift)}

In \cite{GPL3}   we investigated the quasi 1-dimensional problem
of non-interacting point particles of mass $m$ colliding
 elastically with a single heavy ``piston"
of mass $M\gg m$. Initially the piston is fixed at $X=0$ and the particles on the left (resp. right)
of the piston are in thermal equilibrium described by a Maxwellian distribution of velocities (parallel
to the axis of the cylinder\footnote{The other velocity components, 
playing no role in this problem, can be initially
set to zero and then remain zero at all times.})
with temperature $T_0^-$ and uniform density $n_0^-$,
i.e. pressure $p_0^-=n_0^-k_BT_0^-$ (resp.
$T_0^+$, $n_0^+$, $p_0^+$). At time $t_0$ the piston is let free to move
without any friction. The initial conditions ($p_0^{\pm}$,  $T_0^{\pm}$) were chosen such that the piston,
which moves stochastically under the collisions with the particles, remains on the average at $X=0$.
It was then shown that the system evolves to a nonequilibrium
 stationary state (Fig.~1) where the piston has a temperature $T_P$
(average kinetic energy)
and the fluids on the left/right of the piston are characterized 
by temperatures
$T^-\neq T^+$, pressure $p^-=p^+=p$, heat current $j_Q^-=j_Q^+=j_Q$,
and no drift ($w^-=w^+=0$).

% Figure 1
\vspace*{-40mm}\hspace*{40mm}
\leavevmode
\epsfxsize= 60pt
\epsffile[10 15 200  700]{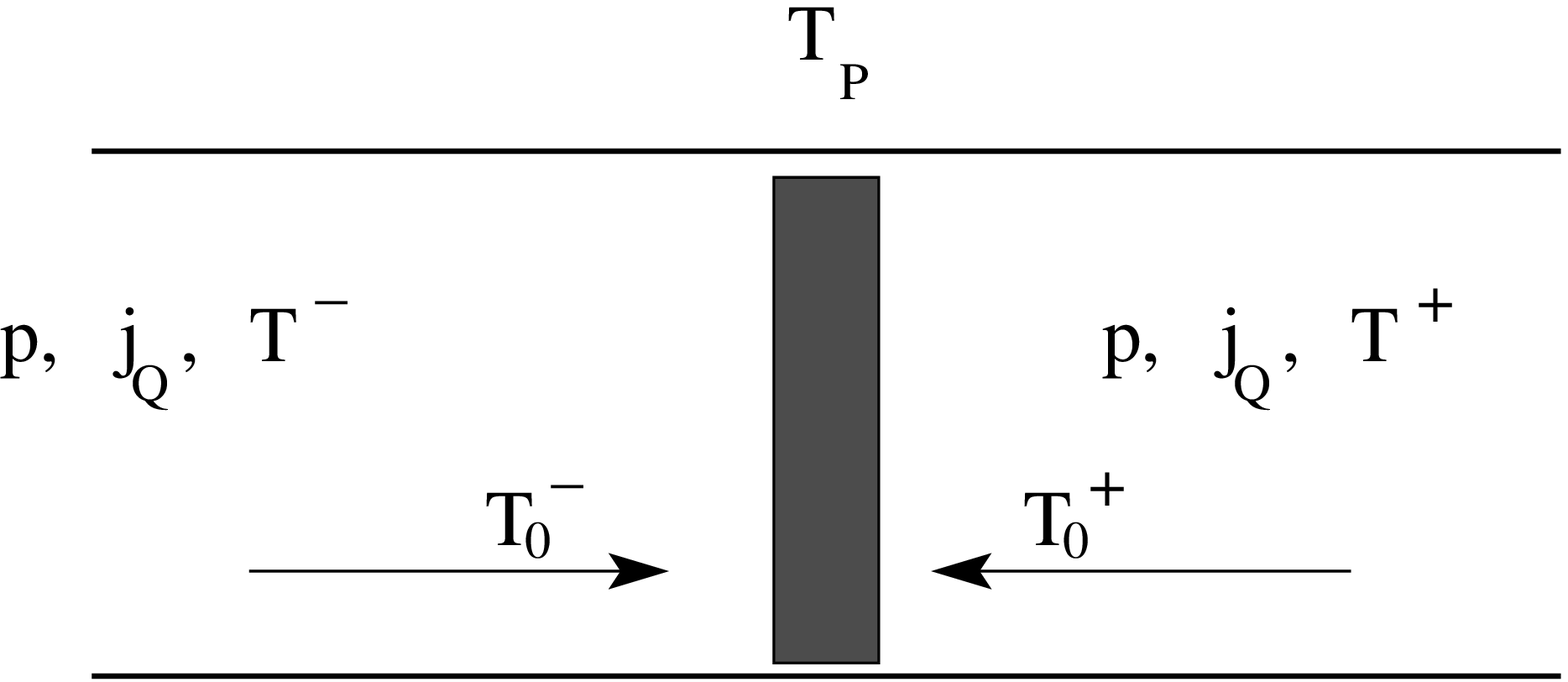}

\vskip 5mm
{\small\bf Figure 1:} {\small\sf Stationary state of the piston problem
(with no drift).
If $T_0^+\neq T_0^-$, 
 it is a nonequilibrium state with $j_Q\neq 0$
and $T^+\neq T^-$.}
% (then $T^+\neq T_0^+$ and $T^-\neq T_0^-$).}
\vskip 5mm

\subsection{Stationary state of the fluid with
prescribed $(p,T,j_Q)$ and no drift }
%Furthermore, i
It was furthermore shown in \cite{GPL3}
 that the nonequilibrium stationary state
 of the fluid parametrized by
\be (p, \;T, \;j_Q,\; w=0)\ee
is characterized by the bimodal distribution function
($\theta$ being the Heaviside function)
%\footnote{
%The connection with \cite{GPL3}
%is  made explicitly by considering
%a two-component fluid with $y=n_1/n_2$}
\be\label{eq:2}
\rho(x,v,t)= \rho(v)={2\beta p\over m}\;
\left\{ \theta[v] {2y\over 1+y} \sqrt{\beta
y\over\pi}\;e^{-\beta y v^2} +\theta[-v] {2\over 1+y} \sqrt{\beta
\over\pi y}\;e^{-\beta v^2/y}\right\}
\ee
where
\be
\beta = {m\over 2k_BT}
\ee
(not to be confused with the inverse temperature $1/k_BT$:
here $\beta^{-1/2}$ is a velocity) and
\be\label{eq:4}
y=1+{C^2\over 2}-{C\over 2}\sqrt{4+C^2}\ee
with $C$  the dimensionless parameter
\be\label{eq:5} C=\sqrt{\pi}\;\beta^{1/2} \;{j_Q\over p}\ee
Let us note that the parameter $y$ is strictly positive. 
Moreover since
\be C={1-y\over \sqrt{y}}\ee
then $j_Q'=-j_Q$ implies $y'=1/y$.
% The  parameters of the fluid nonequilibrium state
%can thus be reformulated  as
%\be (p,\;\beta,\; y=y(p,\beta,j_Q),\;w=0)\ee
%Note that $y<1$ iff $j_Q$ is positive, and $y>1$ iff $j_Q$ is negative.
%For an equilibrium state, $j_Q=0$ implies $y=1$ and we recover the
%usual Maxwellian distribution.
 It is straightforward to
check the consistency between the above result (\ref{eq:2}) for the velocity
distribution and the physical meaning of the associated 
thermodynamic parameters:
\bea
 \int dv \; \rho(v)=n={p\over k_BT}&\hskip 15mm&\mbox{\rm
  (number density $n$ of the ideal fluid)}\\
 &&\nonumber\\
\int dv \;\rho(v)v=0&\hskip 15mm&\mbox{\rm (no drift)}\\
 &&\nonumber\\
 \int dv \;\rho(v)v^2={nk_BT\over m}&
 \hskip 15mm&\mbox{\rm (kinetic temperature $T$ of the fluid)}\label{eq:T}\\
 &&\nonumber\\
 \int dv \;\rho(v)v^3={2\over m}\;j_Q&
 \hskip 15mm&\mbox{\rm (heat flux $j_Q$)}
\eea
It is to be underlined that Eq.~(\ref{eq:2}) reflects the fact that
three different temperatures are involved in the nonequilibrium
stationary state of the fluid (see Fig.~1): the temperature $T_0^-=T/y$ of particles going
to the right ($v>0$), the temperature $T_0^+=Ty$ of particles going
to the left ($v<0$) and the average kinetic temperature $T$
defined by (\ref{eq:T}).
These three temperatures coincide only at thermal equilibrium,
when $y=1$, $j_Q=0$ and $\rho(v)$
coincide with the usual Maxwellian distribution.
They differ as soon as a heat current $j_Q\neq 0$ is forced into the system
and shifts the velocity distribution away from
the equilibrium  Maxwellian distribution.
Note that $y<1$ iff $j_Q$ is positive, and $y>1$ iff $j_Q$ is negative.
The  parameters of the fluid in the nonequilibrium state
can thus be reformulated  as
\be (p,\;\beta,\; y=y(p,\beta,j_Q),\;w=0)\ee

\subsection{Stationary state of the piston}

In the situation sketched in Fig.~1, 
we denote by $T^-$  (resp. $T^+$)
the temperature of the fluid in the stationary state
on the left (resp. right) side of the piston,
and by $T_P$ the temperature of the piston.
In a stationary state,
Eq.~(\ref{eq:2}) is to be written
on each side of the piston with respectively
$\beta=\beta^-=m/2k_BT^-$
and $y^-=y(p, \beta^-, j_Q)$
 in the left compartment,  and
$\beta=\beta^+=m/2k_BT^+$ and $y^+=y(p, \beta^+, j_Q)$
in the right compartment.
%It was shown that (
$X_P$ being  the position of the piston,
it was shown in \cite{GPL3} that in the  stationary state for the piston
\bea
T(X_P-0)=T_0^-\sqrt{1+\delta^-}&
\hskip 10mm&\mbox{\rm (left of the piston)}\label{eq:10}\\
&\hskip 10mm&\nonumber\\
T(X_P+0)=T_0^+\sqrt{1+\delta^+}&\hskip 10mm&\mbox{\rm (right of the piston)}\label{eq:11}\eea
where
\be\label{eq:12}
\delta_{\pm}=\alpha(2-\alpha)\left({T_P\over T_0^{\pm}} -1\right)\ee
with $T_P$ being the temperature of the piston (average
kinetic energy of its stochastic motion) and
\be\label{eq:13}
\alpha={2m\over M+m}\ee

\subsection{Stationarity state  of the piston and surrounding fluids}

Stationarity for the whole system (piston, fluid on the left and fluid on
the right) finally yields the consistency conditions
\be\label{eq:16}
T(X_P-0)=T^-=T_0^-y \hskip 10mm{\rm and}\hskip 10mm
 T(X_P+0)=T^+={T_0^+\over y}\ee
From (\ref{eq:4}) we have
\be\label{eq:16bar}
y-{1\over y}=-\;C\;\sqrt{4+C^2}\ee
and thus from (\ref{eq:11}) and (\ref{eq:12})
\bea
T_{P}
%=T_{P}^L
&=& T^+\left[
y^++{1\over \alpha(2-\alpha)}\left({1\over y^+}-y^+\right)
\right]
\nonumber\\&&
=T^+\;\left[
1+{1\over 2} (C^+)^2 +C^+\sqrt{1+{(C^+)^2\over 4}}\left(
{2\over \alpha(2-\alpha)} -1
\right)
\right]\label{TTPL}
\eea
Similarly, from (\ref{eq:12})
\bea
T_{P}
%=T_{P}^R
&=& T^-\left[
{1\over y^-}+{1\over \alpha(2-\alpha)}\left(y^--{1\over y^-}\right)
\right]
\nonumber\\&&
=T^-\;\left[
1+{1\over 2} (C^-)^2 -C^-\sqrt{1+{(C^-)^2\over 4}}\left(
{2\over \alpha(2-\alpha)} -1
\right)
\right]\label{TTPR}
\eea
From the definition of $C$, Eq.~(\ref{eq:5}), and the fact that 
$p^+=p^-=p$ and  $j_Q^+=j_Q^-=j_Q$, we have 
the equality
$T^+(C^+)^2=T^-(C^-)^2$ relating the fluid nonequilibrium parameters
on each side of the piston in the stationary state;
this relation gives,
using (\ref{TTPL}) and (\ref{TTPR}) 
\be\label{eq:22}
\sqrt{T^+}-\sqrt{T^-}=-\;\left[
{2\over \alpha(2-\alpha)} -1
\right] \;\sqrt{\pi m\over 2k_B}\;
{j_Q\over p}\;\left(1+{\cal O}\left({j_Q\over p}\right)^2\right)\ee
with 
\be\label{eq:20} {2\over \alpha(2-\alpha)} -1 =
{M\over 2m} \;\left(1+{m^2\over M^2}\right)\ee
which shows that
$j_Q={\cal O}(m/M)$, as discussed in \cite{GPL3}. Let us remark that
(\ref{TTPL}) and (\ref{TTPR}) imply also
$$
2T_P=T^++T^-+{\pi m\over 2k_B}\;\left({j_Q\over p}\right)^2+
\hspace{40mm}$$
\be
+\left[
{1\over \alpha(2-\alpha)} -{1\over 2}
\right] \;\sqrt{\pi m\over 2k_B}\;
{j_Q\over p} \;
\left [\;
\sqrt{T^+}\sqrt{4+{\pi m\over 2k_BT^+}\;\left({j_Q\over p}\right)^2}
-\sqrt{T^-}\sqrt{4+{\pi m\over 2k_BT^-}\;\left({j_Q\over p}\right)^2}\;\;
\right]
\ee
Using Eq.~(\ref{eq:22}) and the fact that $j_Q={\cal O}(m/M)$,
we conclude that the temperature of the piston separating the two fluids
is given by
\be\label{eq:23}
T_P=\sqrt{T^+T^-}+{\cal O}\left[\left({m\over M}\right)^2\right]
\ee

\section{Stationary state of two or more  pistons}\label{sect-2piston}
Let us now consider the stationary state defined by $(p, T, j_Q, w=0)$
where the fluid is bounded by two identical stochastic pistons,
each  of mass $M\gg m$,
which remain on the average at the same position under collisions from both
sides (no drift).
% $w=0$).
We shall use the same notations as in \cite{GPL3}
where $T_0^-$ (respectively $T_0^+$) denotes the temperature of
the particles incident on the right piston from the left, i.e.
with $v>0$ (respectively
on the left piston
 from the right, i.e. with $v<0$), and $T$
is the (average kinetic) temperature of the fluid, Eq.~(\ref{eq:T}),
in the intermediary compartment
in the stationary state, see Fig.~2.

% Figure 2
\vspace*{-40mm}\hspace*{40mm}
\leavevmode
\epsfxsize= 60pt
\epsffile[10 15 200  700]{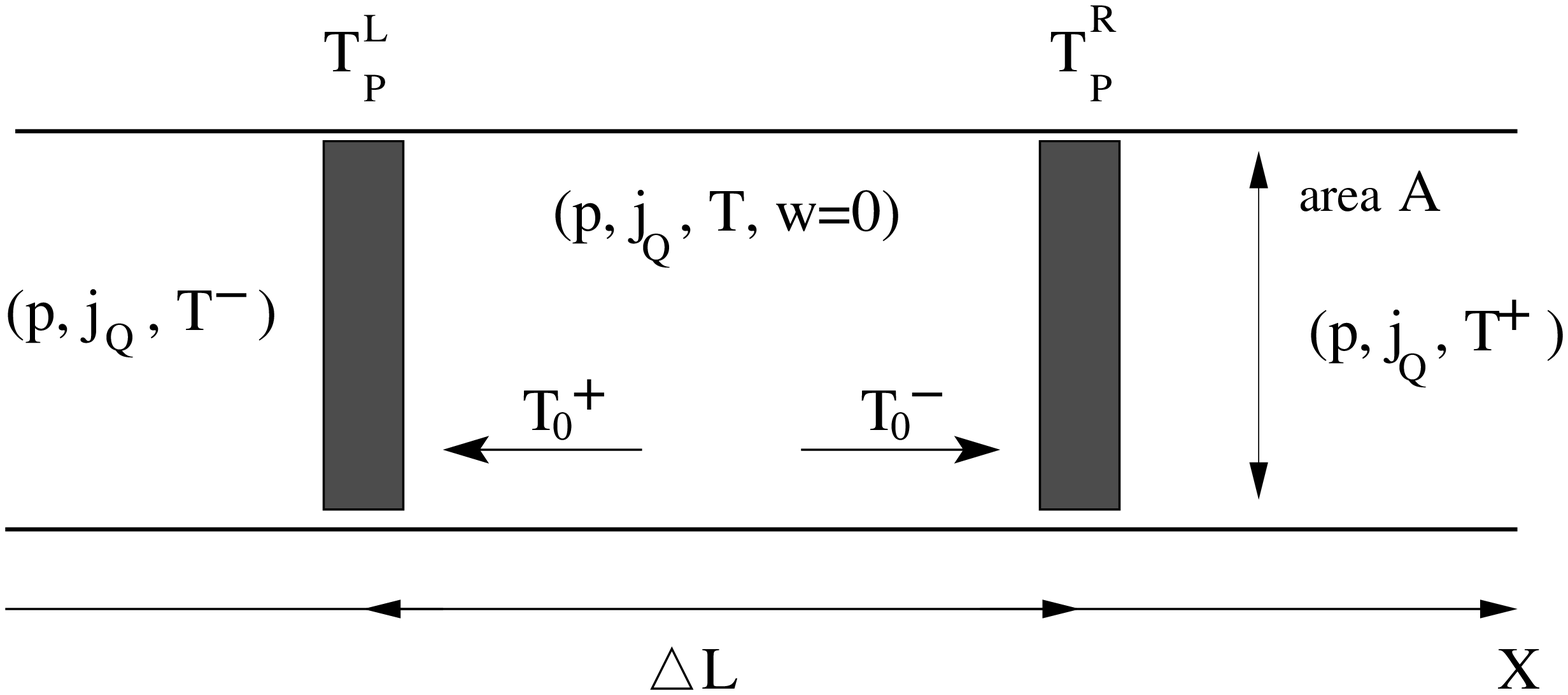}

\vskip 5mm
{\small\bf Figure 2:} {\small\sf  Stationary state with two pistons
with  prescribed current $j_Q$, 
prescribed boundary temperatures $T^{\pm}$,
and no drift (uniform pressure $p$).
$T_0^-$ and $T_0^+$ are auxiliary temperatures, associated
with subpopulations of fluid particles, going respectively
to the right ($v>0$) and to the left ($v<0$) inside the middle
fluid compartment.
}
\vskip 5mm

\subsection{Temperature $T_P^R$ of right piston}
From Eqs. (\ref{eq:10}) and (\ref{eq:16}), we have on the left side 
of the right piston
\be\label{eq:15b}
T_0^-={T\over y}={T\over \sqrt{1+\delta^-}}\ee
(joint stationarity of the fluid and the right piston)
i.e.
\be\label{eq:16b}
y^2=1+\delta^-\ee
Denoting $T_P^R$ the temperature of the right piston (see Fig.~2),
plugging Eqs.~(\ref{eq:15b}-\ref{eq:16b})
in Eq.~(\ref{eq:12}) for $\delta^-$ yields
\be
y^2-1=\alpha(2-\alpha)\left[{T_P^R\over T}y-1\right]\ee
i.e.
\be\label{eq:15}
T_P^R=T\left[{1\over y} +{1\over \alpha(2-\alpha)}\left(y-{1\over y}\right)
\right]\ee

\subsection{Temperature $T_P^L$ of left piston}

Similarly denoting  $T_P^L$ the temperature of the left piston
(see Fig.~2), Eqs. (\ref{eq:11}) and (\ref{eq:16}) give
for the right side of the left piston
\be\label{eq:19b}
T_0^+=T y={T\over \sqrt{1+\delta^+}}\ee
i.e.
\be\label{eq:17}
 (\delta^+)^2={1\over y}-1\ee
and thus from Eqs.(\ref{eq:12}) and (\ref{eq:19b}-\ref{eq:17}) we obtain
\be\label{eq:18}
T_P^L=T\left[y+{1\over \alpha(2-\alpha)}
\left({1\over y}-y\right)
\right]\ee
Let us note that
\be
T_P^R+T_P^L=T\left({1\over y}+y\right)=T(2+C^2)\ee
Therefore the temperature $T$ of the fluid is related
to the temperature of the surrounding pistons by
\be\label{eq:19}
T={1\over 2}(T_P^R+T_P^L)- {\pi m\over 4k_B}
\left({j_Q\over p}\right)^2\ee
Results of Section II can then be applied to each piston
% the
%fluid compartment at temperature $T$ and the outer fluids, at
 to obtain the 
temperature $T^-$ on the left of the left piston and temperature
$T^+$ on the right of the right piston.
Bridging the ensuing relations allows to determine heat conductivity
of this composite system, as follows in the next section.

\section{Expression of heat conductivity}

\subsection{Elementary case of a single fluid
 compartment bounded by two pistons}
 We first derive auxiliary relations for an elementary unit, by considering again
%Let us first consider
 the stationary state for two pistons of area $A$
 separated (on the average) by a distance $\Delta L$, with $\Delta N$
 points particles between them  (see Fig.~2).
From  (\ref{eq:15}) and (\ref{eq:18}), we have
\be
T_P^R-T_P^L=T\;\left({1\over y}-y\right)\;
  \left(1-
{2\over \alpha(2-\alpha)}
\right)\ee
Introducing  Eq.(\ref{eq:16bar}) and 
the expression (\ref{eq:5}) of $C$, together with
(\ref{eq:20}) and $p\;A\;\Delta L=\Delta N\;k_B\;T$, we obtain
\be\label{eq:24}
{1\over \Delta L}\;(T_P^R-T_P^L)= -\; {M\over m} \left(1+{m^2\over M^2}\right)
\sqrt{\pi m\over 2k_B^3}\;
j_Q \left({A\over \Delta N}\right)\;{1\over \sqrt{T}}\;
\sqrt{1+{\pi m \over 8k_BT}
\left({j_Q\over p}\right)^2
}\ee
where the temperature $T$ of the fluid between the two
 pistons is given by (\ref{eq:19}), i.e.
\be\label{eq:25}
T={1\over 2}\;(T_P^R+T_P^L)- {\pi m\over 4k_B}
\left({j_Q\over p}\right)^2\ee
For the fluids on the left and on the right,
Eq.~(\ref{eq:23}) gives at lower order in  $m/M$
\be
T^-={(T_P^L)^2\over T}+{\cal O}\left[\left({m\over M}\right)^2\right]
\hskip 15mm
T^+={(T_P^R)^2\over T}+{\cal O}\left[\left({m\over M}\right)^2\right]\ee

% Figure 3
\vspace*{-40mm}\hspace*{40mm}
\leavevmode
\epsfxsize= 60pt
\epsffile[10 15 200  700]{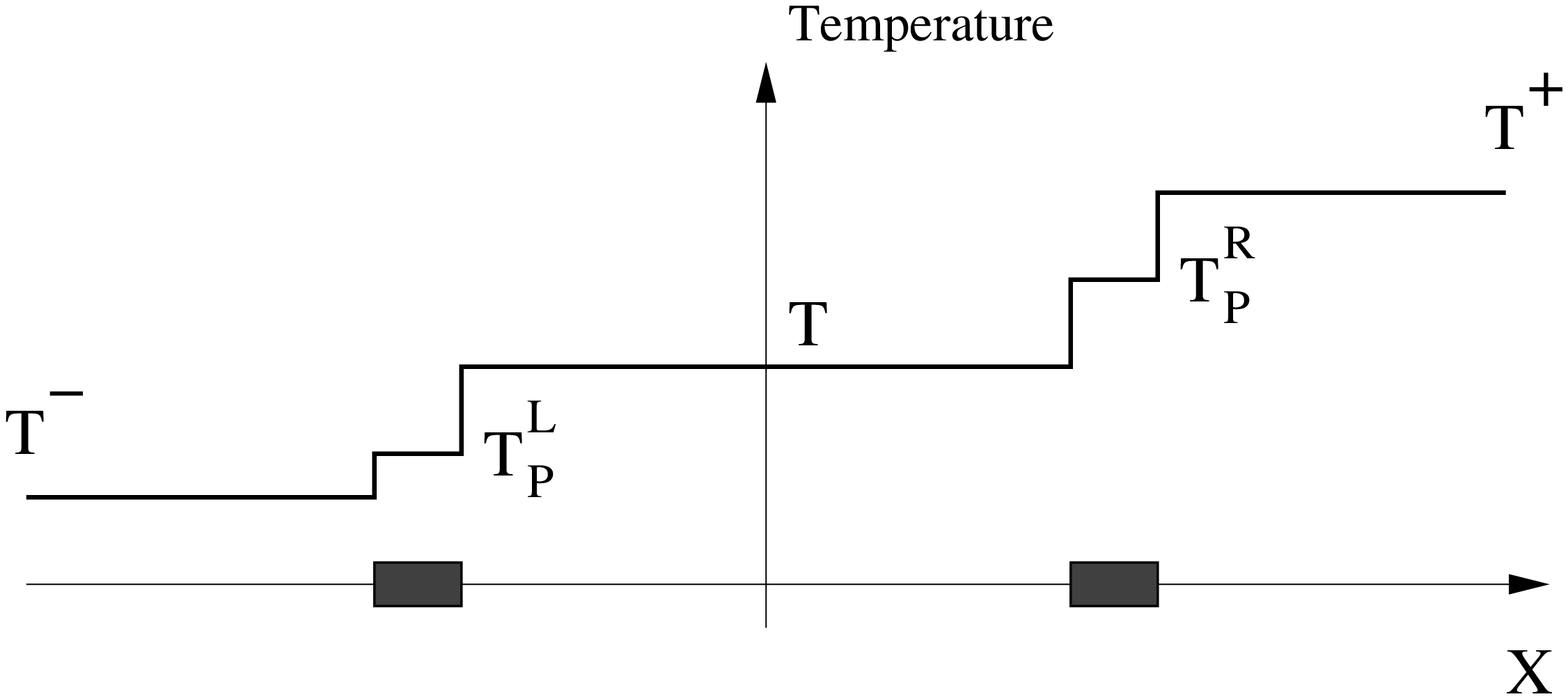}

\vskip 5mm
{\small\bf Figure 3:} {\small\sf  Temperature 
profile for the stationary state of two
identical pistons.
$T_P^L$ and $T_P^R$ are the temperatures (average kinetic energy)
of the pistons, whereas $T^-$, $T$ and $T^+$ are temperatures within
the fluids as defined~in~(\ref{eq:T}).
% and involved as
%parameters in their non-Maxwellian velocity distributions of form (\ref{eq:2}).
}
\vskip 5mm

\subsection{Heat transport through an array of $K$ pistons}
We are now in position  to derive the law describing the
fluctuation-driven heat transport equation for our
model of spatially extended system
and to investigate whether, and in which conditions, the usual Fourier's law
can be recovered.
We now consider the system of $K$
identical (movable) pistons of mass $\Delta M$, with $\Delta N$ point
particles of mass $m$ between them. 
%(which is somehow reminiscent
% of coupled map systems, [refs?]).
 In the stationary state defined
 by $(T^-, p, j_Q)$ on the left,  $(T^+, p, j_Q)$ on the right,
 the pressure and the heat flux in each compartment (defined by successive pistons)
 will also be given by $p$ and $j_Q$, and we denote respectively
 $T_k$ and $T_{P,k}$
 the temperatures of the successive compartments and successive pistons (see Fig.~4).
 The average distance $\Delta L_k$ between adjacent pistons  
 will adapt to
ensure that the pressure is actually homogeneous across the
system (local mechanical equilibrium);
%it comes straighforwardly 
% will be
it is thus  given by
% (see Fig.~4).
\be\label{eq:36}
\langle X_{k+1}\rangle - \langle X_{k}\rangle
=\Delta L_k=\left({\Delta N\over A}\right)\;{k_BT_k\over p}\ee

% Figure 4
\vspace*{-40mm}\hspace*{40mm}
\leavevmode
\epsfxsize= 60pt
\epsffile[10 15 200  700]{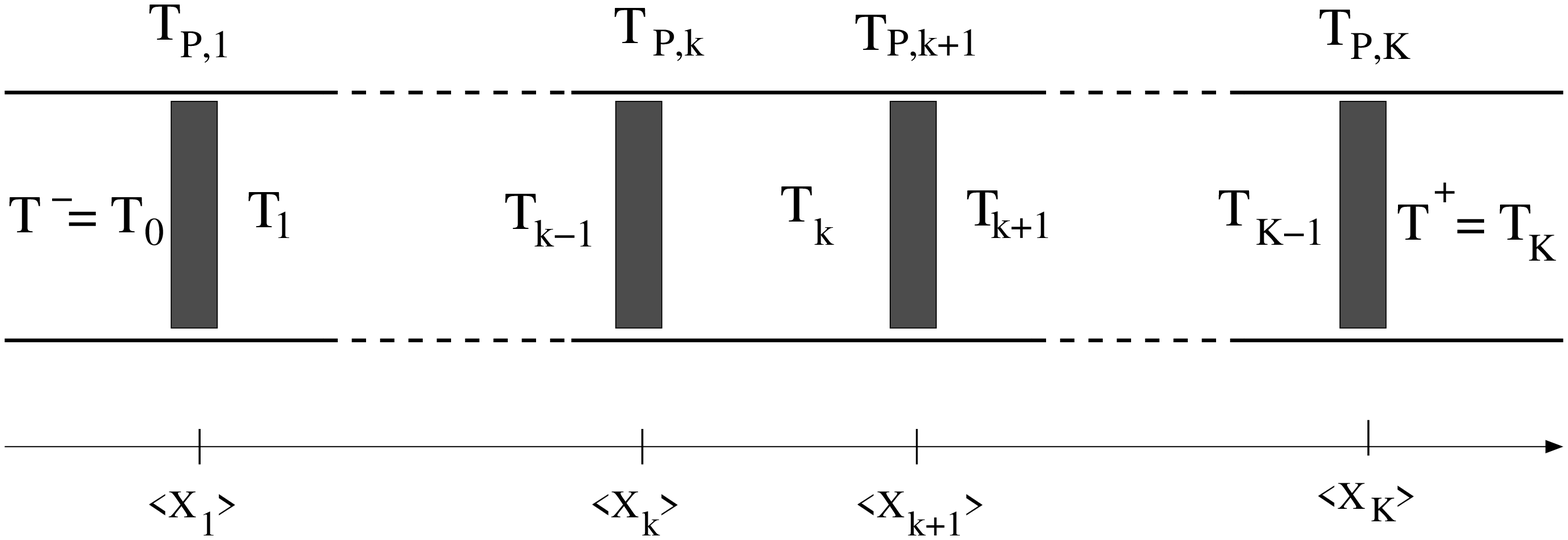}

\vskip 5mm
{\small\bf Figure 4:} {\small\sf Stationary state with $K$
pistons. The pressure $p$ and the heat flux $j_Q$ are identical in
 each subsystem, the drift velocity is zero
and the pistons remain on the average at fixed positions
$\langle X_{k}\rangle$.}
\vskip 5mm

From Eqs.~ (\ref{eq:24}-\ref{eq:25}) applied to the $k$-th compartment
\be\label{eq:28}
{1\over \Delta L_k}(T_{P,k+1}-T_{P, k})=
-\;{\Delta M\over m}\left[
1+\left({m\over \Delta M}\right)^2\right]
\sqrt{\pi m\over 2k_B^3}\left({A\over \Delta N}\right)\;
{j_Q\over \sqrt{T_k}}\;\;
\sqrt{1+{\pi m\over 8k_BT_k}\left({j_Q\over p}\right)^2}
\ee
with
\be \label{eq:29}
T_k={1\over 2}(T_{P,k}+T_{P,k+1}) -{\pi m\over 4k_B}
\left({j_Q\over p}\right)^2\ee
%{\it Vérifier les indices !!!!}
The system considered is defined by the total mass of pistons
$M=K\Delta M$, the total number of point particles
$N=(K-1)\Delta N$, the uniform pressure $p$ and the temperatures
$T^-$ and $T^+$ of the outer (semi-infinite) compartments.
With $K$ sufficiently large, but such that
\be
 \Delta M={M\over K}\gg m, \hskip 15mm
 1\ll \Delta N={N\over K-1}\ll N={ALp\over k_BT}\ee
Eqs.~(\ref{eq:28}-\ref{eq:29}) give
\be\label{eq:30}
\left\{
\begin{array}{l}
\displaystyle
{dT_P(x)\over dx}=-\;{ M\over m}\;{A\over N}\;\sqrt{\pi m\over 2k_B^3}\;
{j_Q\over \sqrt{T}}\;\;
\sqrt{1+{\pi m\over 8k_BT_k}\left({j_Q\over p}\right)^2}
\\
\\
\displaystyle
T(x)=T_P(x)\;\left[1 -
{\pi m\over 4k_BT_P}\left({j_Q\over p}\right)^2
\right]
\end{array}
\right.\ee
where the position of the piston and the location of nearby fluid
can be identified,
and the discrete set of temperature values interpolated
by smooth functions $T_P(x)$ and $T(x)$ 
at this spatial scale and perturbation order.
With $N/A$ given, then $j_Q$ will be of the order $m/M$ in
order that grad $T$ is uniformly bounded. Therefore to first order
in $m/M$, we have
\be\label{eq:31}
\left\{
\begin{array}{l}
\displaystyle
{dT_P(x)\over dx}=-\;{1\over \kappa(x)} \;j_Q\\
\\
\displaystyle
\kappa(x)= {m\over M}\;{N\over A}\;\sqrt{2k_B^3\over \pi m}\;\sqrt{T_P(x)}
\end{array}
\right.\ee
which gives by integration the temperature profile in the system
\be\label{eq:34b}
T_P(x)=T(x)=\left[(T^-)^{3/2} +{x\over L}\left((T^+)^{3/2}-(T^-)^{3/2}\right)
\right]^{2/3}\ee
This temperature profile fulfills, in a purely Hamiltonian setting, the
expected general relation 
\be
T(x)=\left[(T^-)^{\alpha} +{x\over L}\left((T^+)^{\alpha}-(T^-)^{\alpha}\right)
\right]^{1/\alpha}\ee
given in \cite{JPE}, 
%,  with $\alpha=3/2$ as in other instances where the
%energy is purely kinetic,
 hence supports its universality.
 We here find an exponent $\alpha=2/3$  as mentionned in \cite{JPE}, related
 to the fact that the fluids are confined to a limited
 region between pistons and the 
energy is purely kinetic.
%even outside local thermodynamic equilibrium.
Our result moreover shows that local thermodynamic equilibrium
 is not required to get such a temperature profile provided there
is a natural way to define local temperature, here through (\ref{eq:2}),
see also \cite{GPL3}. 
We recover the currently observed
 linear profile if $|T^+-T^-|\ll T^-$, or in  boundary layers
where $x\ll L$ or $(L-x)\ll L$ (there interchanging the role
of $T^+$ and $T^-$  in (\ref{eq:34b})).
%We thus have f
From Eq.~(\ref{eq:31}), we finally get
\be
j_Q=-\kappa\;{dT\over dx}=-\;
{m\over M}\;{N\over A}\;\sqrt{2k_B^3\over \pi m}\;
{2\over 3L}\left[(T^+)^{3/2}-(T^-)^{3/2}\right]\ee
%We here obtain the law in $3/2$-power; as mentionned in \cite{JPE}, related
 %to the fact that the fluids are confined to a limited
%energy is purely kinetic.
Moreover from the knowledge of the integrated
number density  profile $N(x)$ of the fluid, 
such that $N(0)=0$, $N(L)=N$ and (\ref{eq:36})
\be
{dN\over dx}={Ap\over k_BT(x)}=
{Ap\over k_B}\left[(T^-)^{3/2} +{x\over L}\left((T^+)^{3/2}-(T^-)^{3/2}\right)
\right]^{-2/3}\ee
we obtain the length $L$ of the system:
\be
Nk_B=3\,p\;A\;L\;{\sqrt{T^+}-\sqrt{T^-}\over
(T^+)^{3/2}-(T^-)^{3/2}}\ee
i.e.
\be
p\;A\;L\;={Nk_B\over 3}\;
[T^++T^-+\sqrt{T^+T^-}]\ee
Therefore
\be\label{eq:43}
j_Q(x)=-\;
{m\over M}\;\sqrt{8k_B\over \pi m}\;p\;[\sqrt{T^+}-\sqrt{T^-}]\ee
which shows that the term we have neglected when passing  from
(\ref{eq:30}) to (\ref{eq:31})
is of the order $(m/M)^2$.

\section{Fourier's law}

We have thus established that for our model, the Fourier's law
\be
j_Q=-\;\kappa\;{\rm grad}\;T\ee
is valid at some mesoscopic scale allowing to consider
the continuous limit of the discrete microscopic temperature profile, 
 with the heat conductivity $\kappa$ given by
\be \kappa=
{m\over M}\;{N\over A}\;\sqrt{2k_B^3\over \pi m}\;\sqrt{T}\ee
We thus find that $\kappa$ depends on the local temperature in  agreement with the general statement made in \cite{JPE} 
when communicating agents  (here the light particles)
%since here tracers (the light particles)
remain confined (here between the adjacent pistons).
% to their compartments (``locked-in'' tracers).
As one could expect, this heat conductivity is proportional
to the number $N/A$ of gas particles per unit area
and also proportional to the ratio $m/M$,
measuring the efficiency of interaction
between the piston and  gas particles.
It is however independent of the
length $L$ of the system (except for $L$-dependent
corrections to the continuous limit
involved in the derivation of the gradient term).
Let us note that for a strictly 1-dimensional system (for which the area $A$ is zero), one should introduce
 the power of heat $P_Q$ which is transmitted through the system. In this case,
Fourier's law reads
\be
P_Q=-\;\bar{\kappa}\;\;{\rm grad} \;T\ee
and we have
\be
\bar{\kappa}={mN\over M}\;\sqrt{2k_B^3\over \pi m}\;\sqrt{T}\ee
Finally, let us recall that 
Clausius-Maxwell-Boltzman obtained a theoretical expression
for $\kappa$ for gases with $\kappa\sim \sqrt{T}$ independent of
 the gas density \cite{joel}.
%kinetic theory applied to a 1-dimensional
%gas predicts the Fourier law with the same temperature
%dependence
$\kappa\sim\sqrt{T}$ for the  conductivity \cite{joel}.

\section{Entropy production and dissipation}

Since the system considered (defined by the $(K-1)$ elementary units
and $K$ pistons) is in a stationary state, its 
thermodynamic entropy $S$ remains constant.
Thus, in the framework of irreversible thermodynamics, we have
\be
{dS\over dt}= I+P_Q\left({1\over T^-}-{1\over T^+}\right)=0\ee
which gives for the total entropy production per unit time of the system
\be
I=P_Q\left({1\over T^+}-{1\over T^-}\right)\nonumber
=
{Nm\over M}\;\sqrt{2k_B^3\over \pi m}\;{2\over 3L}\;(T^+-T^-)^2\;\;
{T^++T^-+\sqrt{T^+T^-}\over T^+T^-[\sqrt{T^+}-\sqrt{T^-}]}\ee
Therefore $I>0$ as required by the Second Law of thermodynamics.
The same argument remains valid locally where the entropy
density $s(x,t)$ satisfies
\be
{\partial \over \partial t}s(x,t)=i(x,t)-
{\partial \over \partial x}\left({j_Q\over T}\right)\ee
(since there is no drift, hence no extra
entropy current coming from
transport of matter). For the stationary state, it implies \cite{GPL3}
for the source term $i(x)$ associated with irreversibility
\be
i(x)={d\over dx}\left({j_Q\over T}\right)\ee
i.e. (recall that $j_Q$ is constant throughout the system)
\bea
i(x)&=& -\;j_Q\;{1\over T^2}\;{\rm grad}\;T \nonumber\\
&=& \kappa\;\left({{\rm grad}\;T\over T}\right)^2 \nonumber\\
&=& {Nm\over M} {1\over A}\sqrt{2k_B^3\over m}
\left[{2\over 3L}((T^+)^{3/2}-(T^-)^{3/2})
\right]^2{1\over \sqrt{T(x)}}
\eea
Let us note that $I$ is also the total entropy production for the system
plus the two ``reservoirs''.
In conclusion, the irreversibility $i(x)$ is proportional
to the parameter $R=Nm/M$.
Moreover, we see that if $R$ is large, which corresponds to the
 strong damping case \cite{GPL} \cite{gary}, i.e. a high dissipation, the heat conductivity
 is large whereas for $R$ small, which corresponds to the weak damping regime,
 i.e. a weak dissipation, the heat conductivity will be negligible.

\section{Conclusions}
We have considered a quasi-one--dimensional system
composed of $K-1$  ($K\gg 1$)
identical subsystems, in contact at the boundaries with two fluids with
temperatures $T^{\pm}$ and the same pressure $p$.
We have shown that
there exists a nonequilibrium stationary state with heat current $j_Q\neq 0$
given by (\ref{eq:43}) 
% for any given heat current $j_Q\neq 0$, this system
% there exists
%possesses a nonequilibrium stationary state
% with
%heat current $j_Q\neq 0$,
 for which the Fourier's law
$j_Q=-\kappa\;\;{\rm grad}\,T$ is satisfied.
 For this model, the heat conductivity $\kappa$
is proportional to $\sqrt{T}$, as could be expected from kinetic theory. It is
also proportional to the factor $R=Nm/M$,
with $Nm$ the total mass of the light particles and $M$ the  total mass of the
heavy particles (or pistons), as expected from previous work where it
was shown that dissipation (i.e. friction) is proportional to $R$.
It should be remarked that $\kappa$ is of the order of the small 
parameter $\epsilon=m/M$ because the length unit introduced is microscopic.
In this stationary state, the fluid in each subsystem is
characterized by the thermodynamic
parameters $(T(x), p, j_Q, w=0)$
%Temperature is here defined as the (local) average kinetic energy: the system
%is {\it not} at local thermodynamic equilibrium, as can be seen from
%(\ref{eq:2}), where the velocity distribution in the fluid at point $x$
%(so that $\beta=\beta(x)=m/2k_BT(x)$) is {\it not} the Maxwellian
%distribution of temperature $T(x)$.
where $T(x)$ is defined as the local average kinetic energy, see
Eq.~(\ref{eq:T}).
However, it is not a state of local equilibrium since the
velocity distribution function is given by the bimodal expression
Eq.~(\ref{eq:2}) and not by the Maxwell distribution with temperature $T(x)$.
%We nevertheless did not discuss
% the r
Relaxation from an
arbitrary initial condition towards this stationary state with no drift
has not yet been investigated, but is expected to hold as was the case for 
one piston.
%is still an open issue.
%envisioned within this paper, in particular where velocity distributions 
%of the fluids are given by (\ref{eq:2}).
The fluids are moreover characterized by the equation of
state $p=nk_BT$, $e=p/2$. Furthermore,
the total length $L$ of the system will adjust to be given by
$p\;A\;L=(Nk_B/3)\;(T^++T^-+\sqrt{T^+T^-})$. Therefore
introducing an ``average temperature" for the whole system
$T_{Av}=(1/3)\;(T^++T^-+\sqrt{T^+T^-})$, we have the equation
of state for a single component general ideal gas \cite{callen}
 $pV=Nk_BT_{Av}$, $E=Nk_BT_{Av}/2$ (since the number of fluid particles
is much large than the number of pistons).
Finally we have shown that the total entropy production per unit time $I$ of our system
(i.e. $K-1$ subsystems)
as well as the total irreversibility $i(x)$ are strictly positive in agreement
 with the Second Law of thermodynamics.
As mentioned in \cite{GPL3},  the entropy of
the ``heat baths'' will change only due to the heat
flow and there is no internal entropy production.
Therefore  the entropy production for the total system
(i.e. system + heat baths) is simply the quantity $I$ computed above.

We have assumed that the pistons are much heavier than the fluid particles. One may wonder whether our results are more general and remain valid
 if this assumption is dropped. It is an open problem, but clearly
 the mass of the ``piston'' must be different from the mass of
 the fluid particles,
 otherwise we would loose the stochasticity necessary for dissipation.
It is to note that the respective roles of fluid compartments
and pistons are  more symmetric than it might seem: two successive fluid 
compartments are coupled through the fluctuations of the intermediate
piston, and  two successive pistons are coupled through their
interaction (during collisions) with the same intermediate fluid
compartment (also fluctuating and out of equilibrium).
We have here considered ideal fluids, i.e. non-interacting light particles.
In the case of a single piston, simulations with non-interacting
particles or with hard-core particles gave (surprinsingly) similar results
\cite{gary} \cite{GPL},
allowing us to conjecture that
the results given in this paper
would remain valid in the case of hard-core fluid particles.
Behavior for particles interacting through a binary
finite-range potential, for which energy no longer reduces
to kinetic energy, is still unknown \cite{JPE}.

We should stress that we have not limited our discussion to states which
 are close to equilibrium
 ($T^+-T^-$ is arbitrary).
No assumption of linear response has thus been involved in the
derivation of Fourier's law. In agreement with a caveat first underlined
by Van Kampen \cite{vankampen},
 macroscopic linearity of the response is not the reflection
of a linearity of the microscopic response, but rather an emerging
feature following from averaging and cancellation of nonlinear 
microscopic contributions. In the present case,
 it is the perturbation approach and the underlying scale separation
$m/M\ll 1$ which leads to cancellation of nonlinear terms at lowest order.

\vskip 10mm
\noindent
{\bf Acknowledgments:}
We would like to thank J.P. Eckmann for 
communicating us a preprint of his work with L.S. Young
and R. Livi for fruitful discussions.
A.L. acknowledges
 the hospitality of Ecole Polytechnique F\'ed\'erale de Lausanne
where part of this work has been  done.

\end{document}